\documentclass[10pt]{article}

\usepackage{amsmath}
\usepackage{amssymb}

\usepackage{graphicx}

\usepackage{cite}

\usepackage{color}

\usepackage[labelfont=bf,labelsep=period,justification=raggedright]{caption}

\makeatletter
\renewcommand{\@biblabel}[1]{\quad#1.}
\makeatother

\date{}

\newcommand{\bfx}{{\bf x}}

\begin{document}

\begin{flushleft}

{\Large \textbf{Quasi-Neutral theory of epidemic outbreaks} } \\
\bigskip
Oscar A. Pinto $^{1,2}$,
Miguel A. Mu\~noz $^2$
\\
$^1$Departamento de F\'{\i}sica, Instituto de F\'{\i}sica Aplicada,
  Universidad Nacional de San Luis - CONICET, 5700 San
  Luis, Argentina.\\ $^2$ Departamento de Electromagnetismo y
  F{\'\i}sica de la Materia and Instituto de F{\'\i}sica
  Te{\'o}rica y Computacional Carlos I. Facultad de Ciencias,
  Universidad de Granada, 18071 Granada, Spain.
\end{flushleft}
\bigskip
\section*{Abstract}
  Some epidemics have been empirically observed to exhibit outbreaks
  of all possible sizes, i.e., to be scale-free or
  scale-invariant. Different explanations for this finding have been
  put forward; among them there is a model for ``accidental
  pathogens'' which leads to power-law distributed outbreaks without
  apparent need of parameter fine tuning. This model has been claimed
  to be related to self-organized criticality, and its critical
  properties have been conjectured to be related to directed
  percolation. Instead, we show that this is a (quasi) {\it neutral
    model}, analogous to those used in Population Genetics and
  Ecology, with the same critical behavior as the voter-model, i.e.
  the theory of accidental pathogens is a (quasi)-neutral theory. This
  analogy allows us to explain all the system phenomenology, including
  generic scale invariance and the associated scaling exponents, in a
  parsimonious and simple way.

\section*{Introduction}
\label{Intro}

Many natural phenomena such as earthquakes, solar flares, avalanches
of vortices in type II superconductors, or rainfall, to name but a
few, are characterized by outbursts of activity. These are typically
distributed as power-laws of their size, without any apparent need for
fine tuning -- {\it i.e.} they are generically scale invariant--
\cite{SOC,SOC2,SOC3,GG,GG2,Paths}.  This is in contrast to what occurs
in standard criticality, where a control parameter needs to be
carefully tuned to observe scale invariance.  The concept of {\it
  self-organized criticality}, which generated a lot of excitement and
many applications in different fields, was proposed to account for
generic scale invariance, i.e. to explain the ``ability'' of some
systems to self-tune to the neighborhood of a critical point
\cite{SOC, SOC2, SOC3}.

The spreading of some epidemics, such as meningitis in human
populations, has been repeatedly reported to exhibit scale-invariant
traits, including a wide variability of both durations and sizes of
outbreaks. Moreover, the ratio of the variance to the mean of the
distribution of meningitis and measles outbreak sizes have been
empirically found to be very large and to grow rapidly with
population-size \cite{Keeling}. This is the hallmark of anomalously
large fluctuations such as those characteristic of the heavy tails of
power-law distributions \cite{Powerlaws}.  Actually, power-laws have
been proposed to fit the statistics of some epidemics such as measles,
pertussis or mumps in some specific locations as the Faroe islands or
Reykjavik for which accurate long-term epidemiological data are
available \cite{Rhodes, Rhodes2,Rhodes3,Jensen}.  Remarkably, in some
cases, more than four orders of magnitude of scaling have been found
\cite{Keeling}. Other features of scale invariance have been reported
for measles \cite{Keeling} and other infectious child diseases
\cite{Philippe,Philippe2}, rabies and bovine tuberculosis
\cite{Harnos}, or cholera \cite{Cholera}.

At a theoretical level, as pointed out by Rhodes and
Anderson\cite{Rhodes, Rhodes2}, the lack of a characteristic scale
in epidemic outbreaks is reminiscent of earthquakes and their
associated (Guttenberg-Richter) power-law distribution.  Actually,
the presence of scale-invariance in measles, pertussis and others
has been justified in \cite{Jensen} by exploiting the analogies
between simple models for such epidemics and well-known
(self-organized) earthquake models \cite{SOC, SOC2,SOC3}.  In what
follows, we focus now on meningitis, for which a related
self-organized mechanism has been recently proposed
\cite{SJ,SJ2,Stollen,Stollen2,PNAS,Guinea}.

The bacteria responsible for meningitis, {\it Neisseria meningiditis}
or meningococcus, is a human commensal: it is typically harmless and
it is present in up to one fourth of the human population
\cite{Maiden}. Infection is transmitted through close contact with
previously infected individuals. It is noteworthy that killing their
hosts is a highly undesirable outcome for bacteria; therefore it makes
sense that evolution selected for hardly harmful bacteria strains.
Nevertheless, the meningococcus can {\it accidentally} mutate into a
potentially dangerous strain, becoming highly damaging or even lethal
for the host. This is an example of a more general type of
``accidental pathogens'' that innocuously cohabit with the host but
that eventually --even if rarely-- mutate causing symptomatic disease
\cite{Maiden}.

Aimed at modeling accidental pathogens and to shed some light on
the reasons for the emergence of scale invariance in meningococcal
epidemics, Stollenwerk, Jansen and coworkers proposed a simple and
elegant mathematical model
\cite{SJ,SJ2,PNAS,Stollen,Stollen2,Guinea}. The {\bf
Stollenwerk-Jansen (SJ) model} is a variant of the basic
susceptible-infected-recovered-susceptible (SIRS) model.  In the
SIRS, individuals can be in any of the following states:
``susceptible'' (S), ``infected'' (I), or ``recovered'' (R) (also
called ``immune'') \cite{May,Hethcote,Murray}. Perfectly mixed
populations are usually considered, i.e. every individual is
neighbor of any other (which is a mean-field assumption in the
language of Statistical Physics or a ``panmictic'' one in the
Ecology jargon).  The additional key ingredient introduced in the
SJ model is a second, potentially dangerous, strain of infected
individuals labeled $Y$. Dangerous strains appear at a very small
(mutation) rate at every contagion event. The dynamics of $Y$ is
almost identical to that of $I$ except for the fact that at a
certain rate $\epsilon$ they can cause meningococcal disease of
newly infected hosts and eventually kill them, $ Y \rightarrow X$
(see below) \cite{SJ,SJ2,PNAS,Stollen,Stollen2,Guinea}.

By working out explicitly the analytical solution of this
mean-field model as well as performing computer simulations, the
authors above came to the counterintuitive conclusion that the
smaller the value of $\epsilon$ the larger the total amount of
individuals killed on average in a given outbreak \cite{SJ,SJ2}.
This apparent paradox is easily resolved by realizing that the
total number of individuals infected with $Y$ grows upon
decreasing $\epsilon$.  Actually, it has been shown that, whilst
for high pathogenicity $\epsilon$ the distribution of the number
of observed $Y$-cases, $s$, is exponential, in the limit $\epsilon
\rightarrow 0$ it becomes power-law distributed, obeying
\begin{equation}
P(s) \sim s^{-(\tau=3/2)} {\cal{F}}(s/s_c).
\label{scaling}
\end{equation}
where $\cal{F}$ is some scaling function and $s_c$ a maximum
characteristic scale controlled by $1/\epsilon$ \cite{SJ, SJ2}.

Observe that the exponent $ \tau= 3/2$ matches that of a critical
branching process, for which the average population of infected
individuals does not either grow or decay in time but stays constant
\cite{branching}. Also, $ \tau= 3/2$ coincides with the exponent for
the distribution of first return times of an unbiased random-walk
\cite{Gardiner}, which describes generically the scaling of avalanches
in mean-field models.

However, the value $\tau =3/2$ predicted by the SJ model in its
mean-field version, does not necessarily correspond to the best
fit to empirical data \cite{Keeling,Rhodes, Rhodes2}. To
scrutinize the possible origin of such discrepancies, one would
like to go beyond the mean-field/panmictic hypothesis by
considering structured populations in which each individual has a
finite local neighborhood.  In what follows we shall study the SJ
model on populations distributed on regular (Euclidean) lattices
in dimensions $d=2$ and $d=1$. The study of more complex networks
(as small-world networks or networks with communities), aimed at
describing more realistically the net of social contacts, is left
for a future work.

As already pointed out \cite{SJ, SJ2}, the SJ model can be
straightforwardly made spatially explicit. Even if analytical or
numerical calculations have not been performed, it was conjectured in
\cite{SJ,SJ2,Stollen,Stollen2} that in $d$-dimensions the SJ model
should be in the {\it directed percolation} class, i.e. the broad and
robust universality class characterizing phase transitions between an
active and an absorbing state \cite{DP,DP2,AS,AS2,AS3,MD}. In the
present case, the absorbing state would be the $Y$-free state obtained
for sufficiently small $Y$-infection rates, while the active or
endemic phase would correspond to a non-vanishing density of $Y$'s. In
particular, if the critical behavior was indeed that of directed
percolation, then $\tau \approx 1.268 $ for two-dimensional
populations (and $\tau \approx 1.108$ for epidemics propagating in one
dimension) (see \cite{avalanches} and \cite{Zillio}).

A priori, the prediction of directed percolation scaling is
somehow suspicious if the model is indeed self-organized; it has
been shown that in general models of self-organized criticality,
as sandpiles, ricepiles, earthquake models, etc. are {\it not} in
the directed percolation class \cite{FES,FES2,Jabo, Jabo2,Jabo3}.
Furthermore, models of self-organized criticality lacking of any
conservation law (as is the case of the SJ model) have been shown
not to be strictly critical, i.e. they are just approximately
close to critical points \cite{NCSOC}; instead the exact solution
by Stollenwerk and Jansen proves that their model is exactly
critical \cite{SJ, SJ2}. These considerations cast some doubts on
the conjecture of the SJ model being a model of self-organized
criticality \cite{SOC,SOC2,FES,FES2,Jabo, Jabo2,Jabo3}. Therefore,
one is left with the following open questions: what is the SJ
model true critical behavior?, what is the key ingredient why
accidental pathogens -- as described by the SJ model-- originate
scale invariant outbreaks?, does such an ingredient appear in
other epidemics?

The purpose of the present paper is to answer these questions by
analyzing the SJ model in spatially extended systems.

\section*{Methods}

The spatially explicit SJ model is defined as follows. Each site of a
$d$-dimensional lattice is in one of the following states $S, I, Y,
R$, or $X$. The dynamics proceeds at any spatial location $\bfx$ and
any time $t$ according to the following one-site processes:
\begin{itemize}

\item Spontaneous recovery of the benign strain: $I \rightarrow R$, at
  rate $\gamma $.

\item Spontaneous recovery of the dangerous strain: $Y \rightarrow R$,
  at rate $\gamma $.

\item Loss of immunity:  $R \rightarrow S$ at rate $\alpha $.

\item Replacement or recovery of diseased: $X \rightarrow S$, at rate
  $\varphi$,

\end{itemize}

\noindent
and two-site processes:

\begin{itemize}

\item Infection with the benign strain: $S + I
  \rightarrow I + I $, at rate $\beta - \mu $.

\item Infection with the dangerous strain: $S + Y \rightarrow Y + Y$,
  at rate $\beta - \epsilon $.

\item Mutation:   $S + I  \rightarrow Y + I$, at rate $\mu $.

\item Disease: $S + Y \rightarrow X + Y$, at rate $\epsilon$.

\end{itemize}
Some comments on these reaction rules are in order. Accordingly to
\cite{PNAS} it is assumed that being infected with one strain protects
against co-infection with a second one. This assumption, which is
supported by some empirical observations \cite{PNAS}, prevents
transitions as $I + Y \rightarrow Y + Y$ from appearing. $X$
(diseased/dead) individuals are immediately replaced by new
susceptible ones, so that the total population size is kept fixed;
i.e. $\varphi \rightarrow \infty$.  A ``back mutation'' ($S + Y
\rightarrow Y + I$) reaction could be introduced, but for most
purposes its rate is so small that it can be neglected. The dynamics
is easily implemented in computer simulations by using the Gillespie's
algorithm \cite{Gillespie} or a variation of it in a rather standard
way.  In the well-mixed case writing the densities of the different
species as $S$, $I$, $Y$, $R$, and $X$, one readily obtains the
following mean-field or rate equations:
\begin{eqnarray}
\dot{S} &=& \alpha R + \phi X  - \beta S (I + Y) \nonumber \\
\dot{I}  &=& (\beta - \mu) S I - \gamma I \nonumber \\
\dot{Y}  &=& (\beta - \epsilon) S Y  - \gamma Y + \mu S I \nonumber \\
\dot{R}  &=&  \gamma ( I + Y ) - \alpha R \nonumber \\
\dot{X}  &=& \epsilon S Y - \varphi X
\end{eqnarray}
satisfying the constraint $S + I + Y + R + X = 1$.  Stollenwerk
and Jansen \cite{SJ, SJ2} worked out the exact solution of this
set of equations, concluding that it is critical in the limit
$\epsilon \rightarrow 0$, and obtaining the explicit form of
Eq.(\ref{scaling}).

Instead, in spatially explicit models this mean-field approach
breaks down: (i) the densities need to be replaced by
spatio-temporal fields, as $S(\bfx,t)$, $I(\bfx,t)$, $Y(\bfx,t)$,
etc; (ii) new (Laplacian) terms describing the map of
nearest-neighbor contacts appear; (iii) fluctuations become
relevant and noise terms need to be added to account for
demographic fluctuations. A full set of such stochastic Langevin
equations can be derived from the microscopic dynamics by using
standard procedures \cite{Gardiner,GF, GF2} and numerically
investigated \cite{split} (results not shown here).

\section*{Results}

We now report on extensive numerical simulations for the SJ model. We
have chosen the following parameter values: $\alpha=0.3$, $\beta =0.2$
and $\gamma=0.1$, and have tried other different sets to confirm the
robustness of the results.  We consider the mutation rate $\mu$ to be
sufficiently small, such that strains generated by consecutive
mutations do not overlap; i.e. outbreaks finish well before a new
mutation appears. For this reason we take as initial condition for any
outbreak a state with a single mutant of the potentially dangerous
strain $Y$ and effectively fix $\mu=0$ during outbreaks.

\subsection*{Two dimensions}
We consider square lattices of linear size $L=32, 64, 128, 256, 512$,
and $L=1024$, take a random initial condition with only $S$ and $I$
individuals, and run the dynamics with periodic boundary conditions,
keeping $\mu=0$, until a steady state is reached.  For instance, for
$L=128$, the steady state is characterized by $<I>\simeq 0.38$,
$<R>\simeq 0.12$ and $<S>\simeq 0.50$ for the set of parameters above.

Once the system sets into its steady state, we place a $Y$ individual
at the geometrical center of the lattice and study its spreading
\cite{AS,AS2,AS3,MD}. To avoid finite size effects, spreading
experiments are stopped once the $Y$-strain touches the system
boundary, and the described procedure is iterated.  Depending on
system size we ran up to $10^7$ independent realizations.  As
customarily done, we monitor: (i) the epidemics size distribution,
analogous to Eq.(\ref{scaling}), for both $Y$ and $X$; (ii) the
average total number of $Y$ as a function of time $N(t)$; (iii) the
surviving probability $P_s(t)$ that the $Y$ strain is still present in
the system at time $t$, and (iv) the average square radius from the
origin of $Y$-infected individuals, $R^2(t)$.  At criticality, these
quantities are expected to scale algebraically as $N(t) \sim
t^{\eta}$, $P_s(t) \sim t^{-\delta}$, and $R^2(t) \sim t^z$, while
they should show exponential cut-offs in the sub-critical (or
absorbing) phase.

\begin{figure}
\begin{center}
  \includegraphics[height=8.0cm,angle=-0]{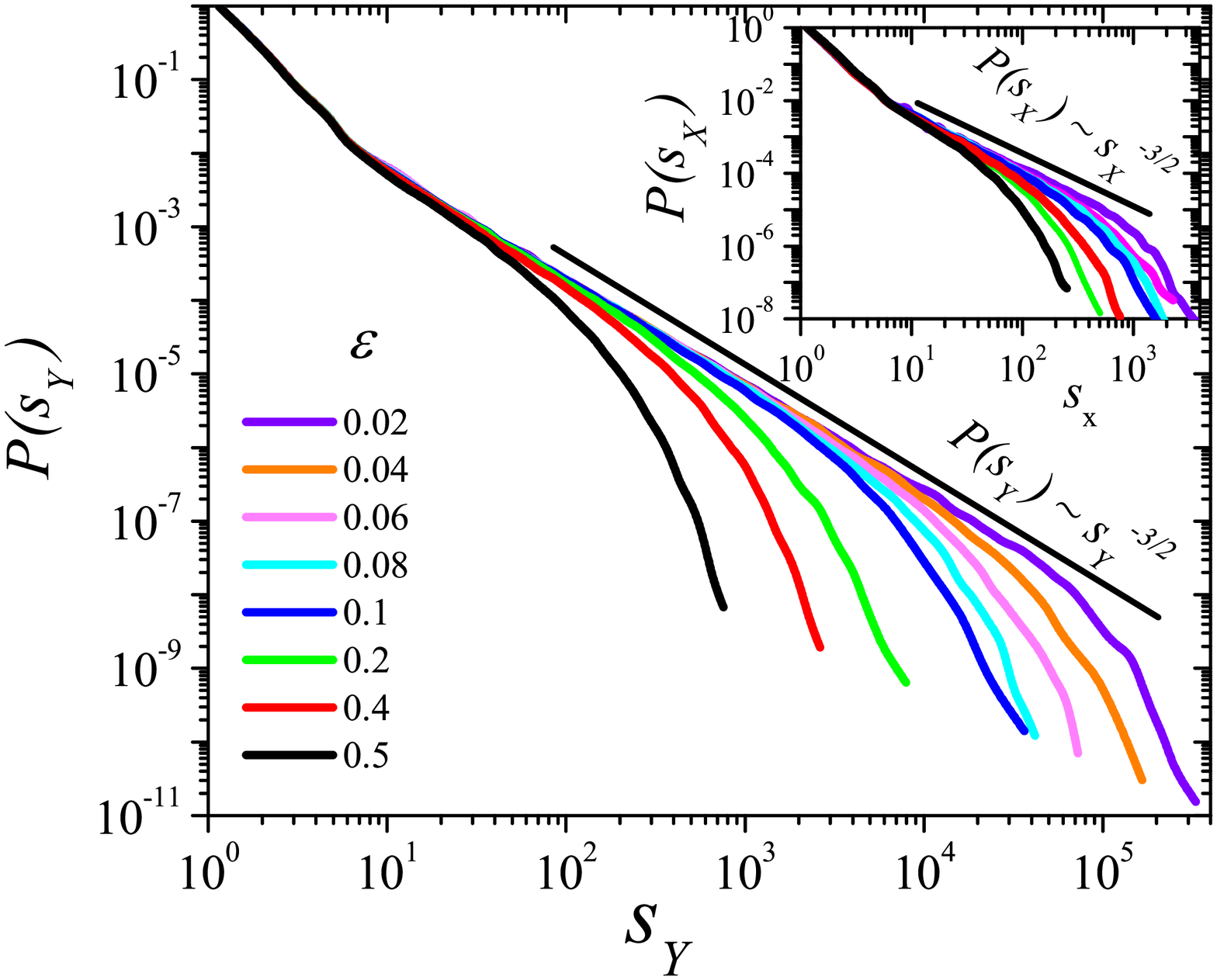}
  \caption{\footnotesize{Main: Avalanche size distribution, $P(s)$ for
      various values of $\epsilon$ in a two dimensional lattice of
      size size $128*128$, and associated distribution for avalanche
      sizes of $Y$-infected individuals, $s_Y$ (main plot), and $s_X$
      of $X$-infected ones (inset).}}
  \label{Ps}
\end{center}
\end{figure}

Fig.\ref{Ps} shows the epidemic outbreak size distribution for
different values of $\epsilon$ and system size $128 * 128$ (size of
$Y$-infected outbreaks, $s_Y$, in the main plot, and size of
$X$-infected outbreaks, $s_x$, in the inset). The probability
distribution of avalanches sizes $s_X$ for $X$-infected sites is
observed to inherit the statistics of $Y$-infected ones. Both
distributions can be well fitted by Eq.(\ref{scaling}) where $\cal{F}$
is a cut-off (exponential) function and $s_c \sim
\epsilon^{-1/\sigma}$ determines the maximum size.  The best fit gives
$\tau \approx 1.48(5)$ (fully compatible with $3/2$ as shown in
Fig.\ref{Ps}) and $1/\sigma \approx 2$ (not shown) both for $P(s_Y)$
and $P(s_X)$. From these plots we conclude that the system becomes
critical in the limit of vanishing $\epsilon$, as occurs in
mean-field.  Observe that for small pathogeneicities, as
$\epsilon=0.02$, the system, even if sub-critical, exhibits scaling
along more than three orders of magnitude.

\begin{figure}
\begin{center}
\includegraphics[height=8.0cm,angle=-0]{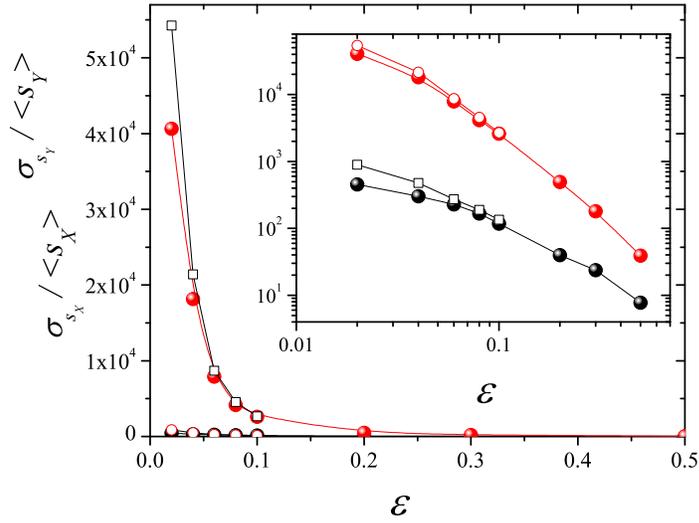}
\caption{\footnotesize{Ratio of the variance ($\sigma_s$) to the mean
    size ($s$) for $Y$ (red) and for $X$ (black), for and different
    values of and system sizes $128*128$ (full symbols) and $256*256$
    (empty symbols). The ratio diverges in the double limit $\epsilon
    \rightarrow 0$, $L \rightarrow \infty$.  Inset: as the main plot,
    but in double logarithmic scale: as system-size increases the
    curves converge to straight lines, i.e. power-laws, but finite
    size effects are significant.}}
  \label{ratioe}
\end{center}
\end{figure}

Both the mean and the variance of the above distributions are observed
to diverge as power-laws in the limit $\epsilon \rightarrow 0^+$.
Actually, as shown in Fig.\ref{ratioe} the ratio of the variance
$\sigma_{s_Y}$ (resp. $\sigma_{s_x}$) to the mean $\langle s_Y\rangle$
(resp. $\langle s_X\rangle$) diverges when $\epsilon \rightarrow 0^+$
as $\sigma_{s_Y} / \langle s_Y \rangle \sim \sigma_{s_X} / \langle s_X
\rangle \sim \epsilon^{-2.2(3)}$ in the infinitely large system-size
limit (otherwise, for finite $L$ a size-induced cut-off appear as
illustrated in Fig.\ref{ratioe} by comparing simulations for two
different sizes).

Analogously, fixing $\epsilon =0$ we measure the mean and variance as
a function of system size; the ratio of the variance to the mean
diverges very fast as $L$ grows, $\sigma_{s_Y} / \langle s_Y \rangle
\sim L^{3.8(3)}$ (results not shown). Observe that, trivially, in this
case $P(s_X)= \delta(s_X)$, i.e. no death is produced.

\begin{figure}
\begin{center}
 \includegraphics[height=8.0cm,angle=-0]{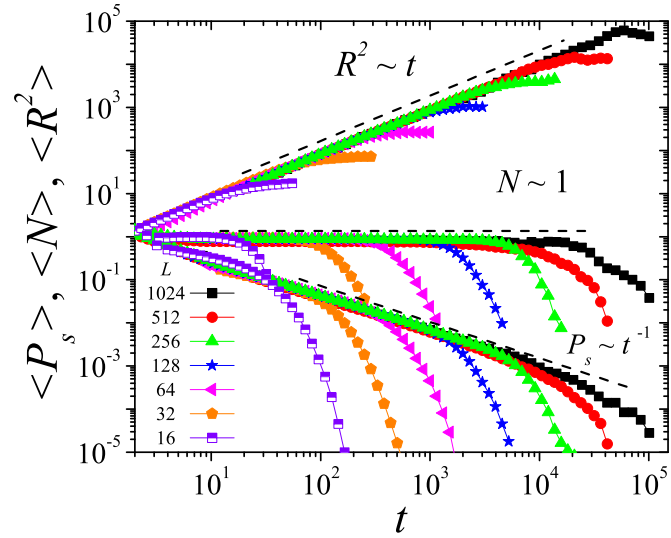}
  \caption{\footnotesize{Log log plot of $N(t)$, $P_s(t)$, and
      $R^2(t)$, for spreading experiments in different two-dimensional
      systems (linear sizes $L=32, 64, 128, 256, 512$, and $1024$) as
      a function of time $t$. Dashed lines are a guide to the eye.}}
  \label{spreading2D}
\end{center}
\end{figure}

\begin{figure}
\begin{center}
  \includegraphics[height=8.0cm,angle=-0]{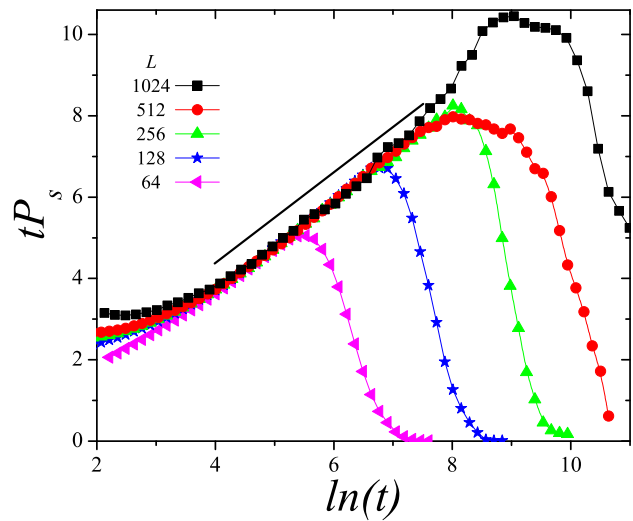}
  \caption{\footnotesize{Plot of $t P_s(t)$ as a function of $\log(t)$
      for different system sizes (from $L=64$ to $L=1024$),
      illustrating the presence of linear logarithmic corrections for
      the surviving probability in the two-dimensional SJ model.}}
  \label{log}
\end{center}
\end{figure}

Fig.\ref{spreading2D} shows results obtained for spreading quantities for
various sizes and $\epsilon=0$, i.e. at criticality.  The best fits we
obtain for the asymptotic behavior of these three magnitudes are
\begin{equation}
 N(t) \sim 1 , ~~~~~ P_s(t) \sim \frac{log(t)}{t}, ~~~~~ R^2(t) \sim t \label{vm}
\end{equation}
An effective power-law, with exponent slightly smaller than unity can
be fit to our numerical results for $P_s(t)$ (see
Fig.\ref{spreading2D}); however, such an effective value of the
exponent can be seen to grow upon extending the maximum time, making
the case for a logarithmic correction as described by Eq.(\ref{vm})
and illustrated in Fig.\ref{log}. Observe also that the reported
values, $\eta=0$, $\delta=1$, $z=1$ satisfy the hyper-scaling relation
$\delta + \eta = dz/2$ using $d=2$ \cite{avalanches}.

Contrarily to a priori expectations and to a previous conjecture,
these asymptotic laws have nothing to do with directed percolation
values (which predicts pure power law behavior for the three
spreading quantities \cite{AS,AS2,AS3,MD} with $\eta \approx 0.229
$, $\delta \approx 0.450 $, $z \approx 1.132 $, $\tau \approx
1.268 $; see \cite{avalanches} for a set of spreading and
avalanche exponent numerical values in different universality
classes).

 Instead, our results for spreading are in excellent agreement with
 the expectations for the two-dimensional voter-model \cite{Voter,
   Voter2,Voter3} universality class (also called ``compact directed
 percolation class'' \cite{CDP,AS, AS2,AS3})
 \cite{MD,Dornic,avalanches} (see below).  Also, our results for the
 size distribution are in excellent agreement with the two-dimensional
 voter class expectations, $\tau=3/2$ and $1/\sigma=2$.
In particular, using these theoretical values, the variance to mean
ratio should scale as $L^4$ in good agreement with the numerical
finding above.

\subsection*{One dimension}

To further confirm the conclusion above, we have also performed
studies of one-dimensional lattices, for which the expected behavior
in the voter-model class is:
\begin{equation}
N(t) \sim t^{0}, ~~~~~
 P_s(t) \sim t^{-1/2}, ~~~~~
R^2(t)  \sim t,
\label{s1d}
\end{equation}
while the avalanche size exponent is $\tau=4/3$ \cite{avalanches}.

\begin{figure}
\begin{center}
  \includegraphics[height=8.0cm,angle=-0]{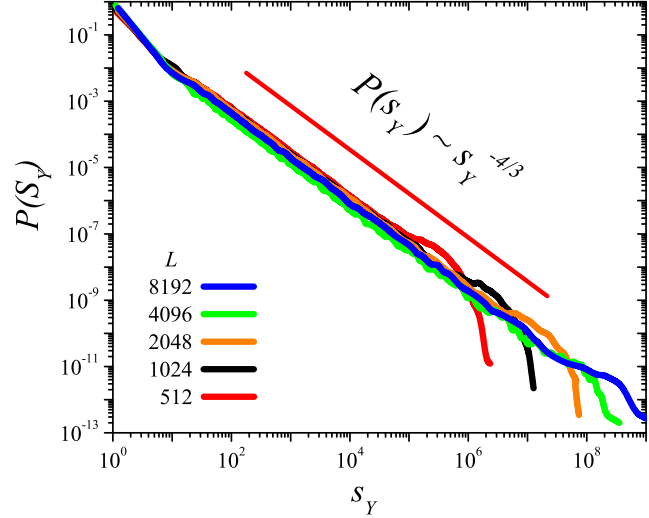}
  \caption{\footnotesize{Avalanche size distribution, $P(s_Y)$
      for $\epsilon=0$ and various one-dimensional lattices, of size
      $L=512, 1024, 2048, 4096$, and $8192$, respectively. The
      straight line corresponds to the theoretical prediction for the
      one-dimensional voter-model class.}}
  \label{Ps1}
\end{center}
\end{figure}

\begin{figure}
\begin{center}
  \includegraphics[height=8.0cm,angle=-0]{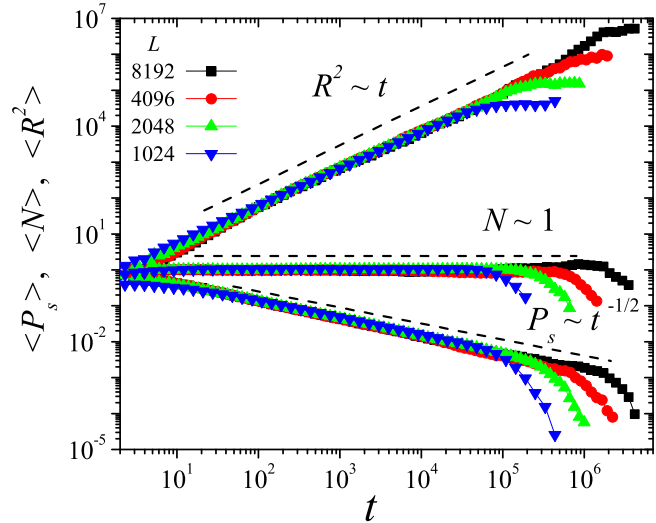}
  \caption{\footnotesize{Evolution of $N(t)$, $P_s(t)$, and $R^2(t)$
        as a function of time $t$ in log-log scale for spreading
        experiments performed on one-dimensional lattices of linear
        sizes $1024, 2048, 4096$, and $8192$.  Parameter values are
        $\alpha=0.3$, $\gamma= 0.1$, and $\beta= 1.8$.  Results are
        averaged over $90000$ independent realizations. }}
  \label{spreading1D}
\end{center}
\end{figure}

Fig.\ref{Ps1} shows the avalanche size distribution for various one
dimensional sizes ($512$, $1024, 2048, 4096$, and $8192$) and
$\epsilon=0$. The measured $\tau$ exponent, $\tau =1.33(3)$ is in
excellent agreement with the voter-model class expectation.
Fig.\ref{spreading1D} shows the result of spreading experiments (for
$L= 1024, 2048, 4096$, and $8192$).  The best fits to the slopes,
$\eta \approx 0$, $\delta =0.50(2)$, and $z=1.01(1)$ are in excellent
agreement with the theoretical prediction (see dashed lines in
Fig.\ref{spreading1D}). These results confirm that the SJ model is in
the voter-class also in one spatial dimension (and exclude
one-dimensional directed percolation values $\eta \approx 0.313 $,
$\delta \approx 0.159 $, $z \approx1.265 $, $\tau \approx 1.108 $).

\section*{Discussion}

Neutral theories date back to the sixties when Kimura introduced them
in the context of population genetics \cite{Kimura}. Kimura assumed,
as a null or neutral model, that each allele of a given gene (in
haploid populations) is equally likely to enter the next generation,
i.e. allele-type does not affect the prospects for survival or
reproduction. Inspired in this, Hubbell proposed an analogous neutral
theory of (forest) bio-diversity, in which the prospects of death and
reproduction do not depend on the tree species \cite{Hubbell}.  Both
theories lead to correct predictions and also to interpretation
controversies (see \cite{Leigh} for a recent review).

In its simplest spatially explicit version the neutral theory is as
follows.  Consider, for argument's sake, the neutral theory of
bio-diversity with only two tree species. It can be formulated by
considering the spatially explicit Moran process \cite{Moran} or {\it
  voter model} \cite{Voter,Voter2, Voter3,MD}. This is defined by two symmetrical
species (up/down, right/left, black/white, etc.) fully occupying a
$d$-dimensional lattice (or, more in general, an arbitrary
network). At some rate a randomly chosen individual is removed and
replaced by any of its nearest neighbors with homogeneous probability.
This leads to a local tendency to create clusters of any of the two
symmetrical species. Clusters of each of the two species will occupy
different positions until eventually one of the two species will take
over the whole (finite) space, leading to fluctuation induced
mono-dominance in any finite system-size,

Such a type of coarsening process has been studied in depth; the
asymptotic scaling of the spreading quantities, $N(t)$, $P_s(t)$, and
$R^2(t)$ is analytically predicted to be given by Eq.(\ref{vm}). In
particular, the logarithmic corrections for the surviving probability
correspond to the fact $d=2$ is the upper critical dimension of the
voter universality class and are closely related to the marginality of
the return time of two-dimensional random walkers \cite{Voter, Voter2,
  Voter3,Zillio}.  The key ingredients of this universality class turn
out to be the symmetry between the two existing absorbing
(mono-dominated) states \cite{AlHammal,DICK,Dornic} and the absence of
surface tension \cite{Dornic}.

It is important to underline that, given the symmetry (neutrality)
between the two species in the voter model, a given population of any
of the species can either grow or decline with the same probability;
i.e. there is no deterministic bias.  Using the field theoretical
jargon, the ``mass'' or ``gap'' term vanishes; all this is tantamount
to the model being critical \cite{Amit}.

Indeed, for the voter model in mean-field, calling $\phi$ the density
in one of the two states (and $1-\phi$ the complementary density for
the second species), then
\begin{equation}
\dot{\phi}= - \phi (1-\phi) + (1-\phi) \phi =0
\label{voter}
\end{equation}
where the first term represents the loss of an individual in the first state
by contact with a neighboring in the second one, and the second
represents the reverse process. Therefore, the average density of
$\phi$ is constant all along the system evolution.  Actually, in the
voter-model there is no control parameter to be tuned; the model lies,
by definition at a critical point: criticality is imposed by the
neutral symmetry. Instead, introducing a bias or preference towards
one of the species,
\begin{equation}
\dot{\phi}= - (1 - \alpha) \phi (1-\phi) +
(1+\alpha) (1-\phi) \phi = \alpha (1-\phi) \phi
\label{voter2}
\end{equation}
where the constant $\alpha>0$ quantifies the degree of asymmetry, a
non-vanishing linear or ``mass'' term is generated, bringing about an
overall tendency for $\phi$ to grow or to decrease depending on the
sign of $\alpha$, i.e. deviating the system from criticality.

Introducing spatial dependence and stochasticity, Eq.(\ref{voter})
transforms into the Langevin equation for the voter class
\cite{Dornic}
 \begin{equation}
\dot{\phi}(\bfx,t)= \nabla^2 \phi(\bfx,t) +
D \sqrt{\phi(\bfx,t)(1-\phi(\bfx,t))}
  \eta(\bfx,t)
\label{Langevin}
\end{equation}
where $\eta(\bfx,t)$ is a Gaussian white noise (observe that the
equation is symmetrical under the change $\phi \leftrightarrow
(1-\phi)$ and that there are two absorbing states at $\phi=0$ and
$\phi=1$, respectively).

Despite of the coincidence in the asymptotic scaling, let us underline
that the SJ model is {\it not} a voter model. In the SJ model there
are $5$ different species and not just $2$. In the case $\epsilon=0$,
however, $I$ and $Y$ are perfectly symmetrical; but as processes as $
Y + I \rightarrow Y + Y$ or $ Y + I \rightarrow I + I$ do not exist,
replacement of one species by the other occurs only if mediated by $S$
particles.  It is, therefore, only at sufficiently coarse grained
scales that the dynamics behaves as the voter-model; microscopically
the two dynamics differ significantly.

Even if the SJ is not a voter model, the underlying reason for it to
exhibit scale-invariance is that, in the limit $\epsilon= 0$ and
$\mu=0$ the model is ``neutral'' (i.e. strains $I$ and $Y$ are
perfectly symmetrical) and, as a direct consequence, it is critical.
More explicitly, calling $Z$ the total density of infected sites
which includes both $I$ and $Y$ then, at a mean field level,
keeping $\mu=0$ and $\epsilon=0$
\begin{eqnarray}
\dot{S} &=& \alpha R  - \beta S Z \nonumber \\
\dot{Z}  &=& \beta  S Z - \gamma Z \nonumber \\
\dot{R}  &=&  \gamma Z - \alpha R,
\end{eqnarray}
and the steady state is $Z_{st}=(\alpha \beta - \alpha \gamma)/(\alpha
\beta + \beta \gamma)$, $S_{st}=\gamma/\beta$, and $R_{st}= Z_{st}
\gamma/\alpha$.  Let us suppose, for argument sake, that $Z_{st} >0$;
i.e. that there is a non-vanishing stationary density of infected
sites (otherwise epidemics would just extinguish in finite time) and
that at time $t=0$ is $Y(t=0) = Y_0$. The evolution of $Y$ is
controlled by
\begin{equation}
\dot{Y}  = \beta  S_{st} Y  - \gamma Y =0,
\label{MFY}
\end{equation}
implying $Y(t)=Y_0$, i.e. there is no bias for $Y$ to grow or decay,
or in other words, under these conditions the model is critical in
what respects the $Y$ variable.  For example, in spreading experiments
we fix $Y_0=1/N$ (and $I_0=Z_{st}-1/N$), implying that the total
number of $Y$-sites is $1$ on average, in agreement with the numerical
findings in Fig.\ref{spreading2D} and Fig.\ref{spreading1D}.
Switching on a non-vanishing $\epsilon$ is equivalent in the voter
model to introduce a bias, inducing deviations from criticality, i.e.
exponential cut-offs for the size distribution and spreading
quantities, as indeed observed in our numerical simulations for
$\epsilon \neq 0$ (see, for instance, Fig.\ref{spreading2D} and
Fig.\ref{spreading1D}).

It is noteworthy that in the perfectly symmetrical (neutral) case,
$\epsilon=0$, $\mu=0$, the model is somehow dull: two benign strains
compete in a critical way, but there is no observable consequence of
this for the population under study.  The interesting behavior of the
SJ model comes from slight deviations from criticality, this is, $\mu
\rightarrow 0$ and $\epsilon \rightarrow 0 $ with $\epsilon >> \mu$;
it is in this double limit that outbreaks leave behind an observable
scale-invariant distribution of sick/dead individuals.

In summary, the Stollenwerk-Janssen model provides us with the most
parsimonious explanation for the appearance of scale-invariant
epidemic outbreaks caused by accidental pathogens such as the
meningococcus. Our main finding is that the system is critical in the
limit of vanishing pathogenicity in all dimensions, and the reason for
this is that the SJ is a {\it neutral model}, which turns out to be
critical for the very same reasons as other neutral theories in
Population Genetics and Ecology are critical. As a consequence of
this, the critical behavior of the model is {\it not} described by
directed percolation as previously conjectured, but instead by the
voter-model universality class, representative of neutral theories.
Understanding the theory of accidental pathogens as a neutral theory
gives us new insight into the origin of generic scale invariance in
epidemics in particular and in propagation phenomena in general.

Similar ideas might apply to related problems as the spreading of
computer viruses in the Internet or in the web of e-mail contacts: the
largest overall damage is expected to occur for any type of spreading
agents if their probability to cause damage is as small as possible.
Being close to neutrality warrants success on the long term.

\vspace{0.5cm}

\section*{Acknowledgments:} We acknowledge J.A. Bonachela, A. Maritan,
and L. Seoane for useful discussions as well as for a critical reading
of early versions of the manuscript.

\newpage


\end{document}